\documentclass[aip,apl,
  reprint,
  amsmath,amssymb]{revtex4-2}

\usepackage{graphicx}
\usepackage{dcolumn}
\usepackage{bm}
\usepackage{xcolor}
\usepackage{float}
\usepackage{xfrac}
\usepackage{verbatim}
\usepackage[section]{placeins}
\usepackage[LGR,T1]{fontenc}

\usepackage[per-mode=reciprocal,uncertainty-mode=separate]{siunitx}
\DeclareSIUnit\bar{bar}
\DeclareSIUnit{\mbar}{\milli\bar}
\DeclareSIUnit\angstrom{\text {Å}}
\DeclareSIUnit{\A}{\angstrom}
\DeclareSIUnit{\invA}{\per\angstrom}

\usepackage[version=4]{mhchem}

\makeatletter
\def\@email#1#2{%
 \endgroup
 \patchcmd{\titleblock@produce}
  {\frontmatter@RRAPformat}
  {\frontmatter@RRAPformat{\produce@RRAP{*#1\href{mailto:#2}{#2}}}\frontmatter@RRAPformat}
  {}{}
}%
\makeatother
\begin{document}

\preprint{APS/123-QED}

\title{Spot Profile Analysis Low Energy Electron Diffraction of Plasma-Enhanced Chemical Vapor Deposition Grown Epitaxial Few-Layer Graphene on Sapphire}
\author{Niels~Ganser$^{1}$, Marko~A.~Kriegel$^{1}$, Umut~Kaya$^{2}$, Jixi~Zhang$^{3}$, Rodney D.~L.~Smith$^{3}$, Marika~Schleberger$^{1,4}$, Wolfgang~Mertin$^{2,4}$, Gerd~Bacher$^{2,4}$,  Michael~Horn-von~Hoegen$^{1,4}$}
\email[Corresponding author:]{michael.horn-von-hoegen@uni-due.de}
\affiliation{$^{1}$Faculty of Physics, University of Duisburg-Essen, Lotharstr.~1, 47057 Duisburg, Germany\\
$^{2}$Werkstoffe der Elektrotechnik, Faculty of Engineering, University of Duisburg-Essen, Bismarckstraße~81, 47057 Duisburg, Germany\\
$^{3}$Department of Chemistry and Waterloo Institute for Nanotechnology, 200~University Ave. W., Waterloo ON N2L 3G1, Canada \\
$^{4}$Center for Nanointegration (CENIDE), Carl-Benz-Straße 199, 47057 Duisburg, Germany\\}

\date{\today}


\begin{abstract}

We demonstrate the use of high-resolution spot-profile analysis low-energy electron diffraction to determine the mean grain size of plasma-enhanced chemical vapor deposition grown few-layer graphene on sapphire (\ce{Al2O3}). The diffraction patterns exhibit broadened graphene spots, pronounced diffuse scattering, and azimuthally extended features, indicating finite crystallite size and rotational disorder. By analyzing the finite-size broadening of the specular (00) spot with an Airy-type diffraction profile, we determine a mean grain diameter of \qty{3.7}{\nm} for the as-grown graphene layer. Post-growth annealing under ultrahigh-vacuum conditions increases the mean grain size to about \qty{5.7}{\nm} and \qty{6.8}{\nm}, respectively. These results establish SPA-LEED as a sensitive reciprocal-space method for quantifying the structural coherence of directly grown graphene on insulating substrates.
\end{abstract}

\keywords{2D materials; Graphene; Sapphire; Plasma enhanced chemical vapor deposition (PECVD); SPA-LEED} 

\maketitle


\section{Introduction} 
The scalable synthesis of two-dimensional materials directly on insulating substrates is a key requirement for their integration into electronic, optoelectronic, and sensor devices~\cite{Abajo2025,Han2025,Yin2025,Yuan2025}. Graphene~\cite{Novoselov2004,Geim2007}, in particular, can be grown with outstanding structural quality by chemical vapor deposition (CVD) on catalytically active transition-metal substrates \cite{Li2009,Coraux2008,Loginova2009,VanGastel2009,Sutter2010,Iwasaki2011,Reddy2011,Hattab2011,VoVan2011,Bu2025,Mohanty2026}. However, transferring graphene from such substrates often introduces contamination, wrinkles, cracks, and other defects \cite{Mattevi2011,Rochford2013,Banszerus2015,Neumaier2019,Liu2024,Guo2026}. Direct growth on technologically relevant insulating substrates, such as sapphire (\ce{Al2O3}) \cite{Kaya2026,Lin2014,Saito2014,Wang2016,Mishra2019,Chen2021,Chang2022,Ai2024} or \ce{SiO2} \cite{McNerny2014,Barbosa2019,RodriguezVillanueva2021,Li2023}, is therefore highly desirable, but remains considerably more challenging because these surfaces lack the catalytic activity required for efficient precursor decomposition and ordered graphene formation. Plasma-enhanced chemical vapor deposition (PECVD) offers a route to overcome this limitation by activating the precursor species in the plasma and thereby enabling graphene growth at reduced substrate temperatures~\cite{Kaya2025,Munoz2018,Mischke2020,Bekduz2020,Lozano2023,Jankauskas2024}. These lower growth temperatures are attractive from a technological perspective, but they also impose severe constraints on the structural quality of the resulting graphene films. In particular, the electrical and transport properties of PECVD-grown graphene are governed by the average grain size, the density of grain boundaries, rotational disorder, and the number of graphene layers \cite{Ferrari2000,Malard2009}. Reliable and quantitative methods for determining these structural parameters are therefore essential for correlating growth conditions with material quality and device performance. 

In this letter, we demonstrate that high-resolution spot-profile analysis low-energy electron diffraction (SPA-LEED) provides a sensitive reciprocal-space method for determining the mean crystallite size of PECVD-grown few-layer graphene on \ce{Al2O3}. The finite lateral size of the individual graphene grains leads to a pronounced broadening of the graphene diffraction spots. By analyzing the broadening of the (00)-spot, we determine a mean grain size of  \qty{3.7}{\nm} for as-grown graphene. Post-growth annealing increases the mean grain size.
These results establish SPA-LEED as a quantitative tool for assessing the structural quality of directly grown graphene on insulating substrates.

\section{Methods and Experimental} 
The graphene layers for investigation were grown on \ce{Al2O3} by PECVD, with the details described elsewhere~\cite{Kaya2025}. In brief, \ce{Al2O3}(0001) wafers were cleaned and subsequently heated to the growth temperature \(T_\mathrm{g}=\qty{670}{\celsius}\). Graphene was deposited for \qty{30}{\min} in a PECVD reactor system (AIXTRON Ltd.) using a \ce{CH4}/\ce{N2} process gas mixture with a flow rate of $5/200\,$sccm at a total pressure of \qty{5}{\mbar}. The plasma was operated at a power of \qty{30}{W}. After growth, the samples were transferred through air for SPA-LEED analysis as described below.

The mean lateral grain size of the graphene was determined by SPA-LEED~\cite{Henzler1982,Scheithauer1986,HornvonHoegen1999,HornvonHoegen1999b}. The graphene/\ce{Al2O3} samples were introduced into the ultrahigh-vacuum chamber through a load-lock system. Prior to the diffraction measurements, the samples were degassed at \(T=\qty{620}{\celsius}\). This temperature is below the graphene growth temperature of \(T_\mathrm{g}=\qty{670}{\celsius}\), ensuring that the structure and morphology of the graphene layers remain unchanged during degassing. Additional post-growth annealing was performed under ultrahigh-vacuum conditions by backside electron beam heating of the \ce{Al2O3} substrate to \(\qty{840}{\celsius}\) and \(\qty{960}{\celsius}\), respectively. All SPA-LEED measurements were carried out at room temperature. The instrumental response function of the SPA-LEED system was determined from a clean \ce{Al2O3} substrate covered with a sub-monolayer amount of carbon to suppress charging during electron diffraction.
All LEED patterns were recorded at an electron energy of \qty{69}{\eV}, where the graphene diffraction spots exhibit maximum intensity.
Raman spectra were acquired as 10 accumulations of \qty{10}{\second} using backscattering geometry on a Renishaw inVia Microscope system. A \qty{532}{\nm} laser (Renishaw DPSSL laser, \qty{50}{\mW}) was filtered to \qty{0.91}{\mW} power at the sample surface with neutral density filters. The use of a \qty{2400}{\per\mm} diffraction grating leads to a spectral resolution of \qty{0.9}{\per\cm}. A 50$\times$ objective, with a numerical aperture of 0.5, was used to focus the laser on the sample and collect the scattered light. The Raman shift was calibrated to a silicon standard, with the peak at \qty{520.5}{\per\cm}. All Raman measurements were carried out in air with the sample at room temperature.

\section{Results and Discussion}

\begin{figure}[t]
  \centering
  \includegraphics[width=0.9\columnwidth]{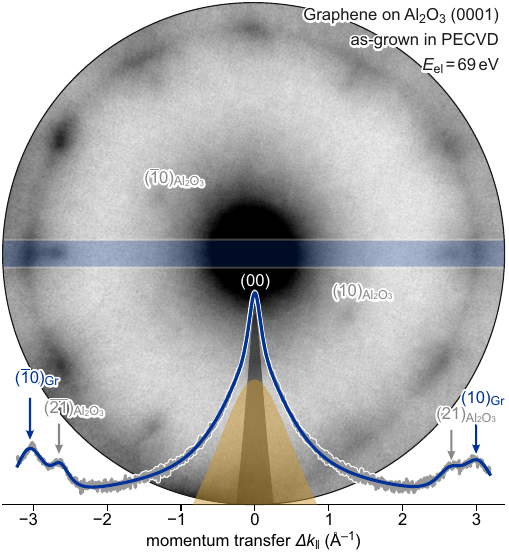}
  \caption{SPA-LEED pattern of the as-grown graphene layer on \ce{Al2O3}(0001), taken at an electron energy of \qty{69}{\eV}. The contrast was enhanced and an inverted gray scale used to emphasize the weak first order diffraction features with the (00) spot overexposed. The inner set of diffraction spots reflects the hexagonal symmetry of the \ce{Al2O3}(0001) substrate. The outer set of spots is assigned to graphene and exhibits an apparent twelvefold symmetry together with pronounced azimuthal and radial broadening. The intensity line profile was taken along the high-symmetry direction indicated by the blue area and displayed in a logarithmic intensity scale. The radially broadened first-order spots at \(k_{\parallel}=\pm\qty{2.641}{\invA}\) and \(k_{\parallel}=\pm \qty{3.03+-0.02}{\invA}\) are attributed to diffraction from \ce{Al2O3}(0001) and graphene, respectively. The broad diffuse intensity underneath the (00) spot, highlighted in yellow, corresponds to the bell-shaped component, commonly observed in LEED from weakly bound two-dimensional materials \cite{Chen2019,Omambac2021,Petrovic2021}. The contribution of graphene to the (00)-spot is marked by the gray shaded peak area.}
  \label{2D-pattern}
\end{figure}

Figure~\ref{2D-pattern} shows a representative SPA-LEED pattern of the as-grown graphene film on \ce{Al2O3} after initial degassing. The diffraction pattern is displayed in an inverted logarithmic intensity scale in order to emphasize the weak features of the first order spots. The overall sixfold symmetry of the pattern reflects the hexagonal symmetry of the underlying \ce{Al2O3}(0001) surface (c-plane). In addition to a broadened specular (00)~spot, the diffraction pattern is dominated by diffuse intensity, while weak and broadened diffraction spots are observed. 

An intensity line profile through the first-order diffraction features is shown at the bottom of Fig.~\ref{2D-pattern}. 
The $k_{\parallel}$-axis was calibrated through the $\pm$(21)-spots of \ce{Al2O3} at \(k_{\parallel}=\pm\qty{2.641}{\invA}\), corresponding to the surface lattice parameter of \(a_{\ce{Al2O3}}=\qty{4.758}{\angstrom}\)~\cite{Yim1974,Suzuki1999}. The outer broad, banana-shaped, diffraction features at \(k_{\parallel}=\pm\qty{3.03+-0.02}{\invA}\) correspond to a real-space lattice parameter of \(a_\mathrm{gr}=\qty{2.39+-0.02}{\A}\), which is close to the lattice constant \qty{2.46}{\angstrom} of graphene \cite{Hattab2011}. These peaks are therefore assigned to first-order graphene diffraction. 

\begin{figure}[t]
  \centering
  \includegraphics[width=0.9\columnwidth]{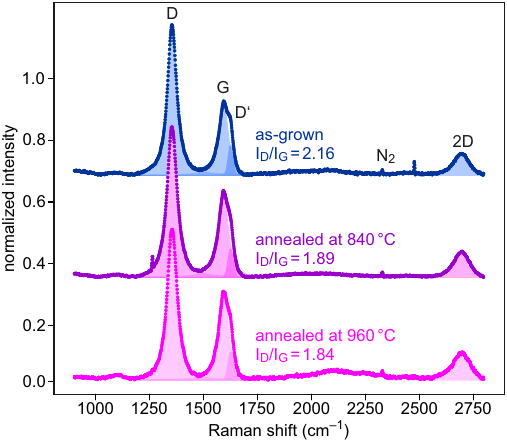}
  \caption{Raman spectra taken for the as-grown graphene sample and after annealing to \(T_{\mathrm{a}}=840\,^{\circ}\mathrm{C}\) and \(T_{\mathrm{a}}=960\,^{\circ}\mathrm{C}\). The large D peaks for all preparations indicate structural disorder and defects. The weak 2D peak is indicative of few-layer graphene. The small feature at \qty{2330}{\per\cm} is the Raman line of \ce{N2} from air~\cite{Rasetti1929}. Spectra are vertically shifted for clarity and were despiked, denoised, baseline-corrected and normalized to the D peak height in RamanSPy~0.2.10~\cite{Georgiev2024}. The Lorentzian fits for the D, G, D', and 2D peak are each indicated as colored areas.}
  \label{Raman}
\end{figure}

Their apparent 12-fold rotational symmetry reflects the  presence of both R0 and R30 oriented epitaxial graphene grains. 
The azimuthal broadening reflects a rotational disorder of graphene grains around the surface normal with a finite rotational spread of approximately $\pm \qty{5+-2}{\degree}$. The radial broadening, on the other hand, is attributed to finite-size effects in diffraction and therefore contains information on the lateral size of the coherently scattering graphene domains. Such finite domain size is supported by Raman spectroscopy, as shown in Fig.~\ref{Raman} for the same as-grown sample. 
The pronounced D peak, with an integrated intensity ratio $I_D/I_G = 2.16$ extracted from the Lorentzian fit, is consistent with a high density of defect- and edge activated scattering sites as expected for nanometer-sized graphene domains. The weak and broadened 2D band is compatible with a few-layer morphology, although its intensity is also strongly affected by disorder.

In addition to the graphene diffraction peaks, a very broad diffuse background with a FWHM $\simeq$~\qty{0.77}{\invA} is observed. We assign this contribution to the so-called bell-shaped component (BSC), reported for several epitaxial 2D material systems \cite{Chen2019,Omambac2021,Petrovic2021}. The BSC is characteristic of LEED from weakly bound 2D layers. It is therefore also observed for the present graphene films, even though the graphene layer consists of small crystalline grains. In these PECVD-grown graphene films this BSC is superimposed on additional finite-size broadening and disorder-related diffuse scattering.

\begin{figure}[ht]
  \centering
  \includegraphics[width=0.9\columnwidth]{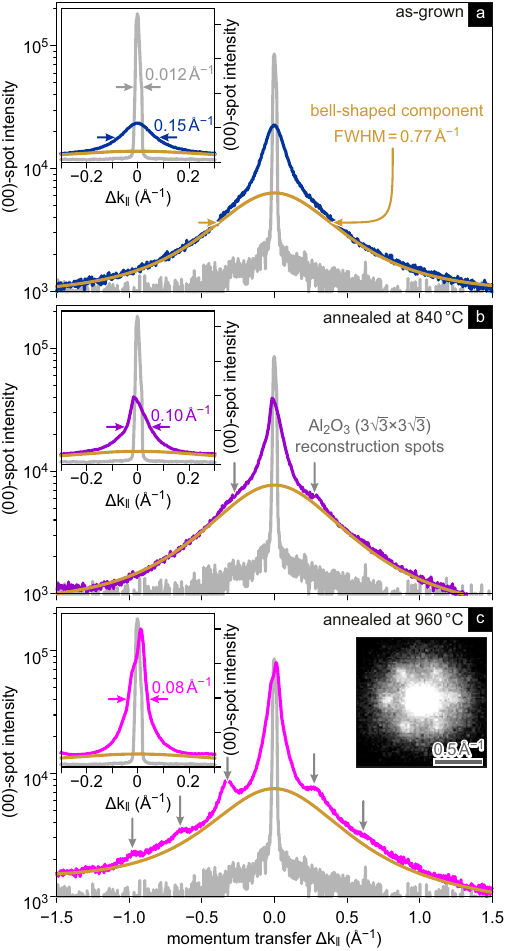}
  \caption{LEED intensity profiles of the (00) spot in logarithmic intensity scale. The insets show a close-up in linear intensity scale. The instrumental response function, recorded as a reference profile from a carbonized \ce{Al2O3}(0001) surface, exhibits a FWHM of the (00) spot of \qty{0.012}{\invA} and is displayed as a gray curve superimposed beneath the spot profiles of the graphene layers. (a) Graphene as grown by PECVD at \(T_\mathrm{g}=\qty{670}{\celsius}\). The broadened (00)~spot exhibits 
  a FWHM of  \qty{0.15}{\invA} 
  It is superimposed on the BSC, shown as a gold solid line, with a FWHM of \qty{0.77}{\invA}. (b,c) Post-growth annealing  at \(T_{\mathrm{a}}=\qty{840}{\celsius}\) and at \(T_{\mathrm{a}}=\qty{960}{\celsius}\) reduces the width of the (00)~spot, while the profile of the BSC remains unchanged. The insets clearly show the increase in intensity and decrease in width of the (00)~spot upon annealing. Grey arrows indicate spots around the (00)-spot which can be attributed to the \ce{Al2O3} substrate. A closeup of the central region is displayed in the inset in (c).}
  \label{Profiles}
\end{figure}

Since the specular (00)~spot is affected by the same finite-size broadening as the higher-order graphene diffraction spots, its profile was used to determine the mean lateral size of the graphene grains. Figure~\ref{Profiles} summarizes this quantitative spot-profile analysis for the different preparation steps. All spot profiles are displayed on a logarithmic intensity scale and over the same reciprocal-space range from \qtyrange{-1.5}{1.5}{\invA}, allowing a direct comparison of the profile shape, background intensity, and spot broadening.

As a reference, the instrumental response function of the SPA-LEED instrument, measured under the same scattering conditions from a single-crystalline \ce{Al2O3}(0001) substrate is superimposed beneath the graphene spot profiles.\FloatBarrier From the FWHM of \(\Delta k_{\mathrm{instr}}=\qty{0.012}{\invA}\), shown more clearly in the linear representation in the inset of Fig.~\ref{Profiles}(a), a transfer width or instrumental coherence length of approximately $2\pi/\Delta k_{\mathrm{instr}}\simeq\qty{50}{nm}$ is obtained \cite{Henzler1982,Scheithauer1986,Comsa1979}. The low diffuse background and the high peak intensity demonstrate the excellent signal-to-noise ratio of the experiment and provide the basis for a reliable spot-profile analysis.

The intensity profile from the as-grown PECVD graphene layer (after degassing) is shown in Fig.~\ref{Profiles}(a) as a solid blue line. In contrast to the instrumental response function, the (00)~spot is strongly broadened and superimposed on the BSC. 
In the logarithmic representation the BSC appears to dominate the profile. The inset of Fig.~\ref{Profiles}(a), however, shows the same spot profile on a linear intensity scale and demonstrates that the finite-width (00) spot can be clearly separated from the BSC. The direct comparison with the instrumental response reveals that the (00)~spot of the graphene-covered \ce{Al2O3}(0001) sample is significantly broader than the instrumental resolution.

For the quantitative analysis, we interpret this broadening as finite-size diffraction from graphene grains of limited lateral extent. Assuming an isotropic distribution of graphene domains, the individual grains are approximated by circular coherently scattering areas. The corresponding diffraction profile is then described by the Airy pattern

\begin{equation*}
    \frac{I(\theta)}{I_0}=\left[\frac{2J_1(x)}{x}\right]^2,
    \qquad x = \frac{\pi D}{\lambda}\sin\theta ,
\end{equation*}

\noindent where $J_1$ is the Bessel function of first order, $D$ is the grain diameter, and $\lambda$ is the electron wavelength~\cite{Hecht2023}. Using this function to calculate the grain diameter from the FWHM measured for the (00)-spot profile in Fig.~\ref{Profiles}(a), after accounting for the instrumental response and the broad diffuse background, yields a mean graphene grain diameter of $D=\qty{3.7}{\nm}$ for the as-grown sample.

The nanometer-scale of the crystalline domains is compatible with the pronounced D band of the Raman measurements in Fig.~\ref{Raman}, which originates from a high density of defect- and edge-activated scattering sites. For a quantitative estimate of the grain size we used the Raman integrated peak intensity ratio of $I_D/I_G = 2.16$. Employing the empirical relations from Tuinstra-Koenig \cite{Tuinstra1970, Matthews1999} and Can{\c{c}}ado~\textit{et.~al.} \cite{Cancado2006} yields values of 2.4~nm and 8.9~nm for the crystallite size, respectively. Because the two Raman-based estimates differ substantially, we refrain from a further quantitative discussion of the Raman-derived grain sizes. Instead, the Raman spectra are used only as qualitative evidence for a high density of defects and finite domain sizes, while the quantitative grain-size analysis is based on the direct reciprocal-space information provided by SPA-LEED.

We further investigated whether post-growth annealing under ultrahigh-vacuum conditions improves the structural quality of the graphene layer. The corresponding spot profiles are shown in Figs.~\ref{Profiles}(b,c). In both cases, the broad BSC remains essentially unchanged and is not affected by the annealing treatment, indicating an intrinsic feature of the graphene layer which supports the interpretation as BSC. Upon annealing, the finite-size-broadened (00)~spot narrows markedly. The FWHM decreases from \(\Delta k=\qty{0.10}{\invA}\) after annealing at \(T_{\mathrm{a}}=\qty{840}{\celsius}\) to \(\Delta k=\qty{0.08}{\invA}\) after \(T_{\mathrm{a}}=\qty{960}{\celsius}\), corresponding to mean grain sizes of \(D=\qty{5.7}{\nm}\) and \(D=\qty{6.8}{\nm}\), respectively. At the same time the spot intensity increases significantly, indicating increasing quality of the graphene film which is consistent with a decrease in relative integrated intensity of the defect-related D peak in Raman spectra, with the $I_D$/$I_G$ ratio decreasing from 2.16 in the as-grown material to 1.89 and 1.84. 

Notably, the first annealing step leads to a clear improvement of the lateral crystalline coherence of the graphene layer, whereas the additional increase in grain size at the higher annealing temperature is comparatively small. This suggests that the enlargement of the graphene grains is limited and that the annealing-induced improvement saturates after the initial structural rearrangement. At the same time, additional diffraction features appear at $\pm \qty{0.31}{\invA}$ and multiples, indicating that further structural changes occur at the graphene/\ce{Al2O3} interface or at the \ce{Al2O3} surface itself as indicated by gray arrows in Fig.~\ref{Profiles}(b,c). These additional spots therefore suggest that excessively high annealing temperatures may no longer improve the graphene layer, but cause dewetting from the substrate, as the additional spots can be explained by the formation of a $(3\sqrt{3} \times 3\sqrt{3})$ surface reconstruction, with spot positions expected at $\pm \qty{0.294}{\invA}$ and which is commonly observed after high temperature treatment~\cite{Chang1968,Gautier1994,Vilfan1997}.
In conclusion, SPA-LEED provides a reliable reciprocal-space method for determining the lateral crystalline coherence and mean grain size of PECVD-grown few-layer graphene on sapphire, even after exposure to ambient conditions. By separating finite-size broadening from the instrumental response and the bell-shaped component, grain sizes in the few-nanometer range can be quantified. This demonstrates SPA-LEED as a sensitive, spatially averaging tool for assessing directly grown two-dimensional materials on insulating substrates.

\section*{Author Contributions}
N.G., M.K., U.K., J.Z. grew the samples, performed the experiments, supporting measurements, and analyzed the data. All authors discussed the results.
N.G. and M.HvH. prepared the figures. 
W.M., G.B., M.S., and M.HvH. conceived and supervised the project. 
The manuscript was written through contributions of N.G., R.S., M.S., and M.HvH.
All authors have given approval to the final version of the manuscript.

\section*{Disclosure of Artificial Intelligence Usage}
ChatGPT-5 was used solely for language editing and for assisting with the formulation of selected passages of the manuscript.

\section*{Declaration of Competing Interest}
The authors declare that they have no known competing financial
interests or personal relationships that could have appeared to influence the work reported in this paper.

\section*{Data availability}
Data is freely available on Mendeley under the following DOI: 10.17632/vrpw9ryv77.1

\section*{Acknowledgements}
Funded by the Deutsche Forschungsgemeinschaft (DFG, German Research Foundation) through project P5 of the International Research Training Group 2803 "Scalable 2D-Materials Architectures - 2D-MATURE" (Project-ID 461605777) and projects B06 and C03 of the Collaborative Research Center SFB1242 "Nonequilibrium dynamics of condensed matter in the time domain" (Project-ID 278162697). We acknowledge software support by F. Thiemann.

\section*{References}
\bibliographystyle{apsrev4-custom}

\bibliography{references}

@article{Abajo2025,
 author = {de Abajo, F. Javier Garc{\'i}a and Basov, D. N. and Koppens, Frank H. L. and Orsini, Lorenzo and Ceccanti, Matteo and Castilla, Sebasti{\'a}n and Cavicchi, Lorenzo and Polini, Marco and Gon{\c{c}}alves, P. A. D. and {Costa {\textit{et.~al.}}}, A. T.},
 year = {2025},
 title = {Roadmap for Photonics with {2D} Materials},
 pages = {3961--4095},
 volume = {12},
 number = {8},
 issn = {2330-4022},
 journal = {ACS photonics}
}

@article{Ai2024,
 author = {Ai, Ding and Yu, Hao and Ma, Yanhao and Cheng, Yonghong and Dong, Chengye},
 year = {2024},
 title = {Modulating growth of graphene on sapphire by chemical vapor deposition},
 pages = {127825},
 volume = {644},
 issn = {00220248},
 journal = {Journal of Crystal Growth}
}

@article{Banszerus2015,
 author = {Banszerus, Luca and Schmitz, Michael and Engels, Stephan and Dauber, Jan and Oellers, Martin and Haupt, Federica and Watanabe, Kenji and Taniguchi, Takashi and Beschoten, Bernd and Stampfer, Christoph},
 year = {2015},
 title = {Ultrahigh-mobility graphene devices from chemical vapor deposition on reusable copper},
 pages = {e1500222},
 volume = {1},
 number = {6},
 journal = {Science advances}
}

@article{Barbosa2019,
 author = {Barbosa, A. N. and Ptak, F. and Mendoza, C. D. and {Da Maia Costa}, M.E.H. and {Freire Jr}, F. L.},
 year = {2019},
 title = {Direct synthesis of bilayer graphene on silicon dioxide substrates},
 pages = {71--76},
 volume = {95},
 issn = {09259635},
 journal = {Diamond and Related Materials}
}

@article{Bekduz2020,
 author = {Bekd{\"u}z, Bilge and Kaya, Umut and Langer, Moritz and Mertin, Wolfgang and Bacher, Gerd},
 year = {2020},
 title = {Direct growth of graphene on {Ge}(100) and {Ge}(110) via thermal and plasma enhanced {CVD}},
 pages = {12938},
 volume = {10},
 number = {1},
 journal = {Scientific reports}
}

@article{Bu2025,
 author = {Bu{\ss}, Lars and Zamborlini, Giovanni and Sulaiman, Cathy and Ewert, Moritz and Cinchetti, Mirko and Falta, Jens and Flege, Jan Ingo},
 year = {2025},
 title = {Hexagons on rectangles: Epitaxial graphene on {Ru} 10\ensuremath{\mathit{\overline{1}}}0},
 pages = {119600},
 volume = {231},
 issn = {00086223},
 journal = {Carbon}
}

@article{Cancado2006,
 author = {Can{\c{c}}ado, L. G. and Takai, K. and Enoki, T. and Endo, M. and Kim, Y. A. and Mizusaki, H. and Jorio, A. and Coelho, L. N. and Magalh{\~a}es-Paniago, R. and Pimenta, M. A.},
 year = {2006},
 title = {General equation for the determination of the crystallite size La of nanographite by Raman spectroscopy},
 pages = {163106},
 volume = {88},
 number = {16},
 issn = {0003-6951},
 journal = {Applied Physics Letters}
}

@article{Chang1968,
 author = {Chang, Chuan C.},
 year = {1968},
 title = {{LEED} Studies of the (0001) Face of \textgreek{a}-Alumina},
 pages = {5570--5573},
 volume = {39},
 number = {12},
 issn = {0021-8979},
 journal = {Journal of Applied Physics}
}

@article{Chang2022,
 author = {Chang, Che-Jia and Tsai, Po-Cheng and Su, Wei-Ya and Huang, Chun-Yuan and Lee, Po-Tsung and Lin, Shih-Yen},
 year = {2022},
 title = {Layered Graphene Growth Directly on Sapphire Substrates for Applications},
 pages = {13128--13133},
 volume = {7},
 number = {15},
 journal = {ACS omega}
}

@article{Chen2019,
 author = {Chen, S. and {Horn von Hoegen}, M. and Thiel, P. A. and Tringides, M. C.},
 year = {2019},
 title = {Diffraction paradox: An unusually broad diffraction background marks high quality graphene},
 pages = {155307},
 volume = {100},
 number = {15},
 issn = {2469-9950},
 journal = {Physical Review B}
}

@article{Chen2021,
 author = {Chen, Zhaolong and Xie, Chunyu and Wang, Wendong and Zhao, Jinpei and Liu, Bingyao and Shan, Jingyuan and Wang, Xueyan and Hong, Min and Lin, Li and Huang, Li and Lin, Xiao and Yang, Shenyuan and Gao, Xuan and Zhang, Yanfeng and Gao, Peng and Novoselov, Kostya S. and Sun, Jingyu and Liu, Zhongfan},
 year = {2021},
 title = {Direct growth of wafer-scale highly oriented graphene on sapphire},
 pages = {eabk0115},
 volume = {7},
 number = {47},
 journal = {Science advances}
}

@article{Comsa1979,
 author = {Comsa, George},
 year = {1979},
 title = {Coherence length and/or transfer width?},
 pages = {57--68},
 volume = {81},
 number = {1},
 issn = {00396028},
 journal = {Surface Science}
}

@article{Coraux2008,
 author = {Coraux, Johann and N'Diaye, Alpha T. and Busse, Carsten and Michely, Thomas},
 year = {2008},
 title = {Structural coherency of graphene on {Ir}(111)},
 pages = {565--570},
 volume = {8},
 number = {2},
 journal = {Nano letters}
}

@article{Ferrari2000,
 author = {Ferrari, A. C. and Robertson, J.},
 year = {2000},
 title = {Interpretation of Raman spectra of disordered and amorphous carbon},
 pages = {14095--14107},
 volume = {61},
 number = {20},
 issn = {2469-9950},
 journal = {Physical Review B}
}

@article{Gautier1994,
 author = {Gautier, Martine and Fenaud, Gilles and {Pham Van}, Laurent and Villette, Bruno and Pollak, Maud and Thromat, Nathalie and Jollet, Fran{\c{c}}ois and Duraud, Jean--Paul},
 year = {1994},
 title = {\textgreek{a}--{Al 2 O 3} (0001) Surfaces: Atomic and Electronic Structure},
 pages = {323--334},
 volume = {77},
 number = {2},
 issn = {0002-7820},
 journal = {Journal of the American Ceramic Society}
}

@article{Geim2007,
 author = {Geim, A. K. and Novoselov, K. S.},
 year = {2007},
 title = {The rise of graphene},
 pages = {183--191},
 volume = {6},
 number = {3},
 journal = {Nature materials}
}

@article{Georgiev2024,
 author = {Georgiev, Dimitar and Pedersen, Simon Vilms and Xie, Ruoxiao and Fern{\'a}ndez-Galiana, {\'A}lvaro and Stevens, Molly M. and Barahona, Mauricio},
 year = {2024},
 title = {RamanSPy: An Open-Source Python Package for Integrative Raman Spectroscopy Data Analysis},
 pages = {8492--8500},
 volume = {96},
 number = {21},
 journal = {Analytical chemistry}
}

@article{Guo2026,
 author = {Guo, Xiaomeng and Qing, Fangzhu and Li, Xuesong},
 year = {2026},
 title = {Correlation Between Crystallinity and Transfer Integrity of {CVD} Graphene},
 pages = {012025},
 volume = {3243},
 number = {1},
 issn = {1742-6588},
 journal = {Journal of Physics: Conference Series}
}

@article{Han2025,
 author = {Han, Jiayue and Fu, Ziyi and Wei, Jingxuan and Han, Song and Deng, Wenjie and Hu, Fangchen and Wang, Zhen and Zhou, Hongxi and Yu, He and Gou, Jun and Wang, Jun},
 year = {2025},
 title = {{2D} materials-based next-generation multidimensional photodetectors},
 pages = {362},
 volume = {14},
 number = {1},
 journal = {Light, science {\&} applications}
}

@article{Hattab2011,
 author = {Hattab, H. and N'Diaye, A. T. and Wall, D. and Jnawali, G. and Coraux, J. and Busse, C. and {van Gastel}, R. and Poelsema, B. and Michely, T. and {Meyer zu Heringdorf}, F.-J. and {Horn-von Hoegen}, M.},
 year = {2011},
 title = {Growth temperature dependent graphene alignment on {Ir}(111)},
 pages = {141903},
 volume = {98},
 number = {14},
 issn = {0003-6951},
 journal = {Applied Physics Letters}
}

@book{Hecht2023,
 author = {Hecht, Eugene},
 year = {2023},
 title = {Optik},
 url = {\url{https://www.degruyterbrill.com/isbn/9783111025599}},
 address = {Berlin and Boston},
 edition = {8., {\"u}berarbeitete Auflage},
 publisher = {{De Gruyter}},
 isbn = {9783111025599},
 series = {De Gruyter Studium},
 institution = {{De Gruyter Oldenbourg}}
}

@article{Henzler1982,
 author = {Henzler, M.},
 year = {1982},
 title = {{LEED} studies of surface imperfections},
 pages = {450--469},
 volume = {11-12},
 issn = {03785963},
 journal = {Applications of Surface Science}
}

@article{HornvonHoegen1999,
 author = {{Horn-von Hoegen}, M.},
 year = {1999},
 title = {Growth of semiconductor layers studied by spot profile analysing low energy electron diffraction -- Part I 1},
 pages = {591--629},
 volume = {214},
 number = {10},
 issn = {2194-4946},
 journal = {Zeitschrift f{\"u}r Kristallographie - Crystalline Materials}
}

@article{HornvonHoegen1999b,
 author = {{Horn-von Hoegen}, Michael},
 year = {1999},
 title = {Growth of semiconductor layers studied by spot profile analysing low energy electron diffraction -- Part II 1},
 pages = {684--721},
 volume = {214},
 number = {11},
 issn = {2194-4946},
 journal = {Zeitschrift f{\"u}r Kristallographie - Crystalline Materials}
}

@article{Iwasaki2011,
 author = {Iwasaki, Takayuki and Park, Hye Jin and Konuma, Mitsuharu and Lee, Dong Su and Smet, Jurgen H. and Starke, Ulrich},
 year = {2011},
 title = {Long-range ordered single-crystal graphene on high-quality heteroepitaxial {Ni} thin films grown on {MgO}(111)},
 pages = {79--84},
 volume = {11},
 number = {1},
 journal = {Nano letters}
}

@article{Jankauskas2024,
 author = {Jankauskas, {\v{S}}arūnas and Me{\v{s}}kinis, {\v{S}}arūnas and {\v{Z}}urauskien{\.{e}}, Nerija and Guobien{\.{e}}, Asta},
 year = {2024},
 title = {Influence of Synthesis Parameters on Structure and Characteristics of the Graphene Grown Using {PECVD} on Sapphire Substrate},
 pages = {1635},
 volume = {14},
 number = {20},
 issn = {2079-4991},
 journal = {Nanomaterials (Basel, Switzerland)}
}

@article{Kaya2025,
 author = {Kaya, Umut and Sahinovic, Armin and L{\"o}rcher, Leon and Nordhoff, Carmen and Zadeh, Yasaman Jarrahi and Lott, Tyler and Sciaini, Germ{\'a}n and Lorke, Axel and Mertin, Wolfgang and Pentcheva, Rossitza and Bacher, Gerd},
 year = {2025},
 title = {{DFT}-Assisted Approach to Low-Temperature Graphene Growth on Sapphire},
 pages = {e07332},
 volume = {21},
 number = {44},
 journal = {Small (Weinheim an der Bergstrasse, Germany)}
}

@article{Kaya2026,
 author = {Kaya, Umut and L{\"o}rcher, Leon and Jarrahizadeh, Yasaman and Lorke, Axel and Mertin, Wolfgang and Bacher, Gerd},
 year = {2026},
 title = {Self--Powered Visible--Blind Graphene/{NiO}/{ZnO} {UV‐C} Photodiodes},
 pages = {e71412, https://doi.org/10.1002/adom.71412},
 issn = {2195-1071},
 journal = {Advanced Optical Materials}
}

@article{Li2009,
 author = {Li, Xuesong and Cai, Weiwei and An, Jinho and Kim, Seyoung and Nah, Junghyo and Yang, Dongxing and Piner, Richard and Velamakanni, Aruna and Jung, Inhwa and Tutuc, Emanuel and Banerjee, Sanjay K. and Colombo, Luigi and Ruoff, Rodney S.},
 year = {2009},
 title = {Large-area synthesis of high-quality and uniform graphene films on copper foils},
 pages = {1312--1314},
 volume = {324},
 number = {5932},
 journal = {Science (New York, N.Y.)}
}

@article{Li2023,
 author = {Li, Mengying and Zou, Pinggen and Chen, Zhi and Zhao, Ruotong and Tang, Cao and Wang, Shuai and Ma, Yanqing and Ma, Lei},
 year = {2023},
 title = {Wafer-Scale Graphene Growth on {Si}/{SiO} 2 Substrates via Metal-Free Chemical Vapor Deposition},
 pages = {10817--10825},
 volume = {6},
 number = {12},
 issn = {2574-0970},
 journal = {ACS Applied Nano Materials}
}

@article{Lin2014,
 author = {Lin, Meng-Yu and Su, Chen-Fung and Lee, Si-Chen and Lin, Shih-Yen},
 year = {2014},
 title = {The growth mechanisms of graphene directly on sapphire substrates by using the chemical vapor deposition},
 pages = {223510},
 volume = {115},
 number = {22},
 issn = {0021-8979},
 journal = {Journal of Applied Physics}
}

@article{Liu2024,
 author = {Liu, Anhan and Zhang, Xiaowei and Liu, Ziyu and Li, Yuning and Peng, Xueyang and Li, Xin and Qin, Yue and Hu, Chen and Qiu, Yanqing and Jiang, Han and Wang, Yang and Li, Yifan and Tang, Jun and Liu, Jun and Guo, Hao and Deng, Tao and Peng, Songang and Tian, He and Ren, Tian-Ling},
 year = {2024},
 title = {The Roadmap of 2D Materials and Devices Toward Chips},
 pages = {119},
 volume = {16},
 number = {1},
 journal = {Nano-micro letters}
}

@article{Loginova2009,
 author = {Loginova, Elena and Nie, Shu and Th{\"u}rmer, Konrad and Bartelt, Norman C. and McCarty, Kevin F.},
 year = {2009},
 title = {Defects of graphene on Ir(111): Rotational domains and ridges},
 pages = {085430},
 volume = {80},
 number = {8},
 issn = {2469-9950},
 journal = {Physical Review B}
}

@article{Lozano2023,
 author = {Lozano, Miguel Sinusia and Bernat-Montoya, Ignacio and Angelova, Todora Ivanova and Mojena, Alberto Bosc{\'a} and D{\'i}az-Fern{\'a}ndez, Francisco J. and Kovylina, Miroslavna and Mart{\'i}nez, Alejandro and Cienfuegos, Elena Pinilla and G{\'o}mez, V{\'i}ctor J.},
 year = {2023},
 title = {Plasma-Induced Surface Modification of Sapphire and Its Influence on Graphene Grown by Plasma-Enhanced Chemical Vapour Deposition},
 pages = {1952},
 volume = {13},
 number = {13},
 issn = {2079-4991},
 journal = {Nanomaterials (Basel, Switzerland)}
}

@article{Malard2009,
 author = {Malard, L. M. and Pimenta, M. A. and Dresselhaus, G. and Dresselhaus, M. S.},
 year = {2009},
 title = {{Raman} spectroscopy in graphene},
 pages = {51--87},
 volume = {473},
 number = {5-6},
 issn = {03701573},
 journal = {Physics Reports}
}

@article{Mattevi2011,
 author = {Mattevi, Cecilia and Kim, Hokwon and Chhowalla, Manish},
 year = {2011},
 title = {A review of chemical vapour deposition of graphene on copper},
 pages = {3324--3334},
 volume = {21},
 number = {10},
 issn = {0959-9428},
 journal = {J. Mater. Chem.}
}

@article{Matthews1999,
 author = {Matthews, M. J. and Pimenta, M. A. and Dresselhaus, G. and Dresselhaus, M. S. and Endo, M.},
 year = {1999},
 title = {Origin of dispersive effects of the {Raman D} band in carbon materials},
 pages = {R6585-R6588},
 volume = {59},
 number = {10},
 issn = {2469-9950},
 journal = {Physical Review B}
}

@article{McNerny2014,
 author = {McNerny, Daniel Q. and Viswanath, B. and Copic, Davor and Laye, Fabrice R. and Prohoda, Christophor and Brieland-Shoultz, Anna C. and Polsen, Erik S. and Dee, Nicholas T. and Veerasamy, Vijayen S. and Hart, A. John},
 year = {2014},
 title = {Direct fabrication of graphene on {SiO2} enabled by thin film stress engineering},
 pages = {5049},
 volume = {4},
 journal = {Scientific reports}
}

@article{Mischke2020,
 author = {Mischke, Jan and Pennings, Joel and Weisenseel, Erik and Kerger, Philipp and Rohwerder, Michael and Mertin, Wolfgang and Bacher, Gerd},
 year = {2020},
 title = {Direct growth of graphene on {GaN} via plasma-enhanced chemical vapor deposition under {N 2} atmosphere},
 pages = {035019},
 volume = {7},
 number = {3},
 journal = {2D Materials}
}

@article{Mishra2019,
 author = {Mishra, Neeraj and Forti, Stiven and Fabbri, Filippo and Martini, Leonardo and McAleese, Clifford and Conran, Ben R. and Whelan, Patrick R. and Shivayogimath, Abhay and Jessen, Bjarke S. and Bu{\ss}, Lars and Falta, Jens and Aliaj, Ilirjan and Roddaro, Stefano and Flege, Jan I. and B{\o}ggild, Peter and Teo, Kenneth B. K. and Coletti, Camilla},
 year = {2019},
 title = {Wafer-Scale Synthesis of Graphene on Sapphire: Toward Fab-Compatible Graphene},
 pages = {e1904906},
 volume = {15},
 number = {50},
 journal = {Small (Weinheim an der Bergstrasse, Germany)}
}

@article{Mohanty2026,
 author = {Mohanty, Smruti Ranjan and Brendel, Lothar and Kriegel, Marko and Ganser, Niels and {Meyer Zu Heringdorf}, Frank-Joachim and {Horn-von Hoegen}, Michael},
 year = {2026},
 title = {Subsurface Carbon Incorporation Controls Graphene Nucleation on Ir(111)},
 pages = {38284--38290},
 volume = {18},
 number = {27},
 journal = {ACS applied materials {\&} interfaces}
}

@article{Munoz2018,
 author = {Mu{\~n}oz, Roberto and Mart{\'i}nez, Lidia and L{\'o}pez-Elvira, Elena and Munuera, Carmen and Huttel, Yves and Garc{\'i}a-Hern{\'a}ndez, Mar},
 year = {2018},
 title = {Direct synthesis of graphene on silicon oxide by low temperature plasma enhanced chemical vapor deposition},
 pages = {12779--12787},
 volume = {10},
 number = {26},
 issn = {2040-3364},
 journal = {Nanoscale}
}

@article{Neumaier2019,
 author = {Neumaier, Daniel and Pindl, Stephan and Lemme, Max C.},
 year = {2019},
 title = {Integrating graphene into semiconductor fabrication lines},
 pages = {525--529},
 volume = {18},
 number = {6},
 journal = {Nature materials}
}

@article{Novoselov2004,
 author = {Novoselov, K. S. and Geim, A. K. and Morozov, S. V. and Jiang, D. and Zhang, Y. and Dubonos, S. V. and Grigorieva, I. V. and Firsov, A. A.},
 year = {2004},
 title = {Electric field effect in atomically thin carbon films},
 pages = {666--669},
 volume = {306},
 number = {5696},
 journal = {Science (New York, N.Y.)}
}

@article{Omambac2021,
 author = {Omambac, K. and Kriegel, M. and Brand, C. and Finke, B. and Kremeyer, L. and Hattab, H. and Janoschka, D. and Dreher, P. and {Meyer zu Heringdorf}, F.-J. and {Momeni Pakdehi}, D. and Pierz, K. and Schumacher, H. W. and Petrovi{\'c}, M. and {van Houselt}, A. and Poelsema, B. and Tringides, M. C. and {Horn-von Hoegen}, M.},
 year = {2021},
 title = {Non-conventional bell-shaped diffuse scattering in low-energy electron diffraction from high-quality epitaxial 2D-materials},
 pages = {241902},
 volume = {118},
 number = {24},
 issn = {0003-6951},
 journal = {Applied Physics Letters}
}

@article{Petrovic2021,
 author = {Petrovi{\'c}, Marin and {Meyer Zu Heringdorf}, Frank-J. and Hoegen, Michael Horn-von and Thiel, Patricia A. and Tringides, Michael C.},
 year = {2021},
 title = {Broad background in electron diffraction of {2D} materials as a signature of their superior quality},
 pages = {505706},
 volume = {32},
 number = {50},
 journal = {Nanotechnology}
}

@article{Rasetti1929,
 author = {Rasetti, F.},
 year = {1929},
 title = {ON THE RAMAN EFFECT IN DIATOMIC GASES},
 pages = {234--237},
 volume = {15},
 number = {3},
 issn = {0027-8424},
 journal = {Proceedings of the National Academy of Sciences of the United States of America}
}

@article{Reddy2011,
 author = {Reddy, Kongara M. and Gledhill, Andrew D. and Chen, Chun-Hu and Drexler, Julie M. and Padture, Nitin P.},
 year = {2011},
 title = {High quality, transferrable graphene grown on single crystal {Cu}(111) thin films on basal-plane sapphire},
 pages = {113117},
 volume = {98},
 number = {11},
 issn = {0003-6951},
 journal = {Applied Physics Letters}
}

@article{Rochford2013,
 author = {Rochford, Caitlin and Kumar, Nardeep and Liu, Jianwei and Zhao, Hui and Wu, Judy},
 year = {2013},
 title = {All-optical technique to correlate defect structure and carrier transport in transferred graphene films},
 pages = {7176--7180},
 volume = {5},
 number = {15},
 journal = {ACS applied materials {\&} interfaces}
}

@article{RodriguezVillanueva2021,
 author = {Rodr{\'i}guez-Villanueva, Sandra and Mendoza, Frank and Instan, Alvaro A. and Katiyar, Ram S. and Weiner, Brad R. and Morell, Gerardo},
 year = {2021},
 title = {Graphene Growth Directly on {SiO2}/{Si} by Hot Filament Chemical Vapor Deposition},
 pages = {109},
 volume = {12},
 number = {1},
 issn = {2079-4991},
 journal = {Nanomaterials (Basel, Switzerland)}
}

@article{Saito2014,
 author = {Saito, Kosuke and Ogino, Toshio},
 year = {2014},
 title = {Direct Growth of Graphene Films on Sapphire (0001) and (11\ensuremath{\mathit{\overline{20}}}) Surfaces by Self-Catalytic Chemical Vapor Deposition},
 pages = {5523--5529},
 volume = {118},
 number = {10},
 issn = {1932-7447},
 journal = {The Journal of Physical Chemistry C}
}

@article{Scheithauer1986,
 author = {Scheithauer, U. and Meyer, G. and Henzler, M.},
 year = {1986},
 title = {A new {LEED} instrument for quantitative spot profile analysis},
 pages = {441--451},
 volume = {178},
 number = {1-3},
 issn = {00396028},
 journal = {Surface Science}
}

@article{Sutter2010,
 author = {Sutter, P. W. and Albrecht, P. M. and Sutter, E. A.},
 year = {2010},
 title = {Graphene growth on epitaxial {Ru} thin films on sapphire},
 pages = {213101},
 volume = {97},
 number = {21},
 issn = {0003-6951},
 journal = {Applied Physics Letters}
}

@article{Suzuki1999,
 author = {Suzuki, T. and Hishita, S. and Oyoshi, K. and Souda, R.},
 year = {1999},
 title = {Structure of \textgreek{a}-{Al 2 O 3} (0001) surface and {Ti} deposited on \textgreek{a}-{Al 2 O 3} (0001) substrate},
 pages = {289--298},
 volume = {437},
 number = {3},
 issn = {00396028},
 journal = {Surface Science}
}

@article{Tuinstra1970,
 author = {Tuinstra, F. and Koenig, J. L.},
 year = {1970},
 title = {Raman Spectrum of Graphite},
 pages = {1126--1130},
 volume = {53},
 number = {3},
 issn = {0021-9606},
 journal = {The Journal of Chemical Physics}
}

@article{vanGastel2009,
 author = {{van Gastel}, R. and N'Diaye, A. T. and Wall, D. and Coraux, J. and Busse, C. and Buckanie, N. M. and {Meyer zu Heringdorf}, F.-J. and {Horn von Hoegen}, M. and Michely, T. and Poelsema, B.},
 year = {2009},
 title = {Selecting a single orientation for millimeter sized graphene sheets},
 pages = {121901},
 volume = {95},
 number = {12},
 issn = {0003-6951},
 journal = {Applied Physics Letters}
}

@article{Vilfan1997,
 author = {Vilfan, Igor and Lan{\c{c}}on, Fr{\'e}d{\'e}ric and Villain, Jacques},
 year = {1997},
 title = {Rotational reconstruction of sapphire (0001)},
 pages = {62--68},
 volume = {392},
 number = {1-3},
 issn = {00396028},
 journal = {Surface Science}
}

@article{VoVan2011,
 author = {Vo-Van, Chi and Kimouche, Amina and Reserbat-Plantey, Antoine and Fruchart, Olivier and Bayle-Guillemaud, Pascale and Bendiab, Nedjma and Coraux, Johann},
 year = {2011},
 title = {Epitaxial graphene prepared by chemical vapor deposition on single crystal thin iridium films on sapphire},
 pages = {181903},
 volume = {98},
 number = {18},
 issn = {0003-6951},
 journal = {Applied Physics Letters}
}

@article{Wang2016,
 author = {Wang, Huaping and Yu, Gui},
 year = {2016},
 title = {Direct {CVD} Graphene Growth on Semiconductors and Dielectrics for Transfer-Free Device Fabrication},
 pages = {4956--4975},
 volume = {28},
 number = {25},
 journal = {Advanced materials (Deerfield Beach, Fla.)}
}

@article{Yim1974,
 author = {Yim, W. M. and Paff, R. J.},
 year = {1974},
 title = {Thermal expansion of {AlN}, sapphire, and silicon},
 pages = {1456--1457},
 volume = {45},
 number = {3},
 issn = {0021-8979},
 journal = {Journal of Applied Physics}
}

@article{Yin2025,
 author = {Yin, Lei and Wang, Junyong and Sun, Yinghui and Liu, Xingqiang and Zou, Xuming and Cheng, Ruiqing and Wang, Hao and Zhu, Yushan and Gong, Xunguo and {Liu {\textit{et.~al.}}}, Zijia},
 year = {2025},
 title = {Recent progress in growth and applications of {2D} materials},
 pages = {100059},
 volume = {1},
 number = {3},
 issn = {30509130},
 journal = {Review of Materials Research}
}


\end{document}